\documentclass[11pt,letterpaper,english,final,aps,prc,tightenlines,superscriptaddress]{revtex4}
\usepackage[T1]{fontenc}
\usepackage[latin1]{inputenc}
\usepackage{amsmath}
\usepackage{graphicx}
\usepackage{amssymb}

\makeatletter

\usepackage{geometry}

\usepackage{array}

\usepackage{longtable}

\usepackage{float}

\makeatletter

\usepackage{amsfonts}

\makeatother

\makeatother

\usepackage{babel}

\begin{document}

\title{Lepton-nucleon interactions at the next-to-the-leading order}

\author{A. Aleksejevs}

\affiliation{ Division of Science, SWGC, Memorial University, Corner Brook, NL, A2H 6P9, Canada}

\author{S. Barkanova}

\affiliation{Department of Physics, Acadia University,  Wolfville, NS, B4P 2B2, Canada}

\begin{abstract}
Next-to-Leading-Order (NLO) effects play a crucial role in tests of the Standard Model (SM), and require careful theoretical evaluation. Electroweak physics, which has just entered the precision age, is an excellent place to search for new physics, but also requires considerable theoretical input. We show how we applied computational packages such as FeynArts, FormCalc, Form and LoopTools for the evaluation of one-loop electroweak and hadronic radiative corrections.  
\end{abstract}
\maketitle

\section{Introduction}

With recent advances in the automatization of the (NLO) calculations, it is reasonable to consider these methods in the 
applications towards the electro-weak and hadronic processes. Computer
packages such as FeynArts \cite{FA}, FormCalc, LoopTools \cite{FCLT}
and Form \cite{Form} created a possibility to perform such type of
calculations. In the work presented here, we extend FeynArts for the
NLO symbolic calculations of amplitude or differential cross section.
Using Dirac and Pauli type couplings, we construct the computational
model \cite{WCharge} enabling us to deal with electron-nucleon scattering
up to NLO level. Using this model we compute parity-violating asymmetries
up to NLO level in electron-proton (e-p) scattering. Additionally we
have developed an extension named Computational Hadronic Model (CHM)
\cite{CHM} of the FeynArts towards the hadronic sector using the
Chiral Perturbation Theory (ChPTh). Later we have applied CHM to extract
strong coupling constants arising in the chiral lagrangian from the
experimental branching ratios (BR) for the following processes: $\Sigma^{*+}\rightarrow\Sigma^{0}+\pi^{+},$
$\Sigma^{*-}\rightarrow\Lambda+\pi^{+},$ $\Xi^{*0}\rightarrow\Xi^{*-}+\pi^{+}$
and $\Delta^{++}\rightarrow p+\pi^{+}$.

\section{Electroweak Sector}

Electroweak properties of the nucleon can be studied by parity-violating
electron-nucleon scattering at low energies. Such experiments can
measure the asymmetry coming from the difference between cross sections
of left- and right-handed electrons ($A=\frac{d\sigma_{+}-d\sigma_{-}}{d\sigma_{+}+d\sigma_{-}}$).
This asymmetry between left- and right-handed particles is clearly
predicted in the Standard Model (SM) of particle physics. Extracting
the physics of interest from the measured asymmetry requires evaluating
NLO contribution to electroweak scattering at very high precision.
The dominant contribution normally comes from the Leading Order (LO)
correction in perturbation theory. The NLO effects in the physics
of the electroweak interactions plays a crucial role in the tests
of the Standard Model. In this project, we took into consideration
the NLO effects in parity-violating lepton scattering and have computed
radiative corrections to the parity violating asymmetries. In general,
we have extended FeynArts by including Dirac and Pauli form factors
in couplings taken in the dipole/monopole form without strange quark
contribution. Calculations were done in the on-shell renormalization
scheme using Feynman gauge. Detailed description of this model is
given in \cite{WCharge}. To avoid the infrared divergences, we have
treated the final asymmetries with both Soft and Hard Photon Bremsstrahlung
(SPB+HPB) \cite{ABB2006} contributions. Computed results for the
asymmetry are given in the Fig.{[}1a{]} for the range of the momentum
transfers up to $1.0\, GeV^{2}$ in the forward scattering. 

\begin{figure}
\begin{centering}
\includegraphics[scale=0.23]{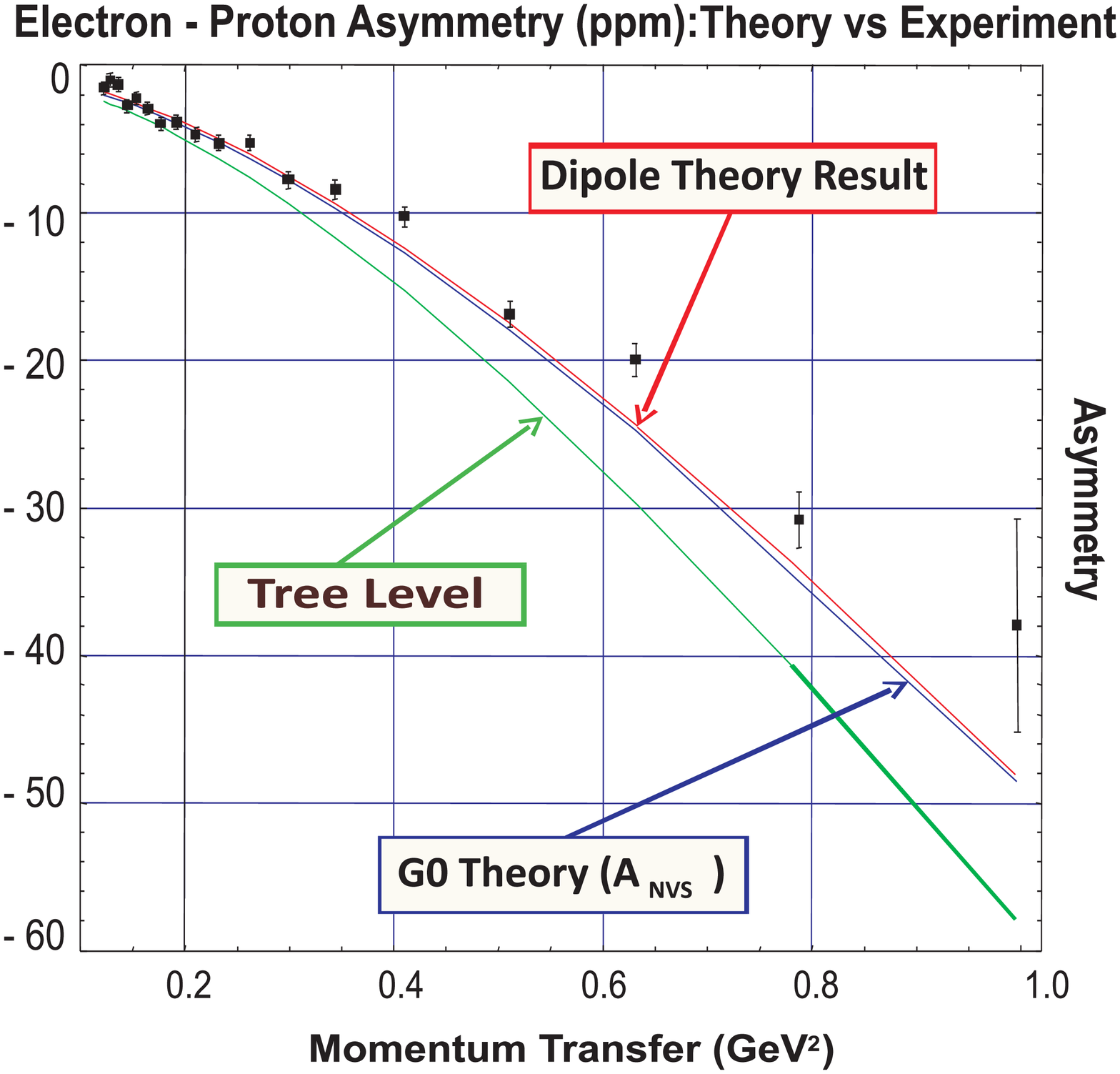}\includegraphics[scale=0.23]{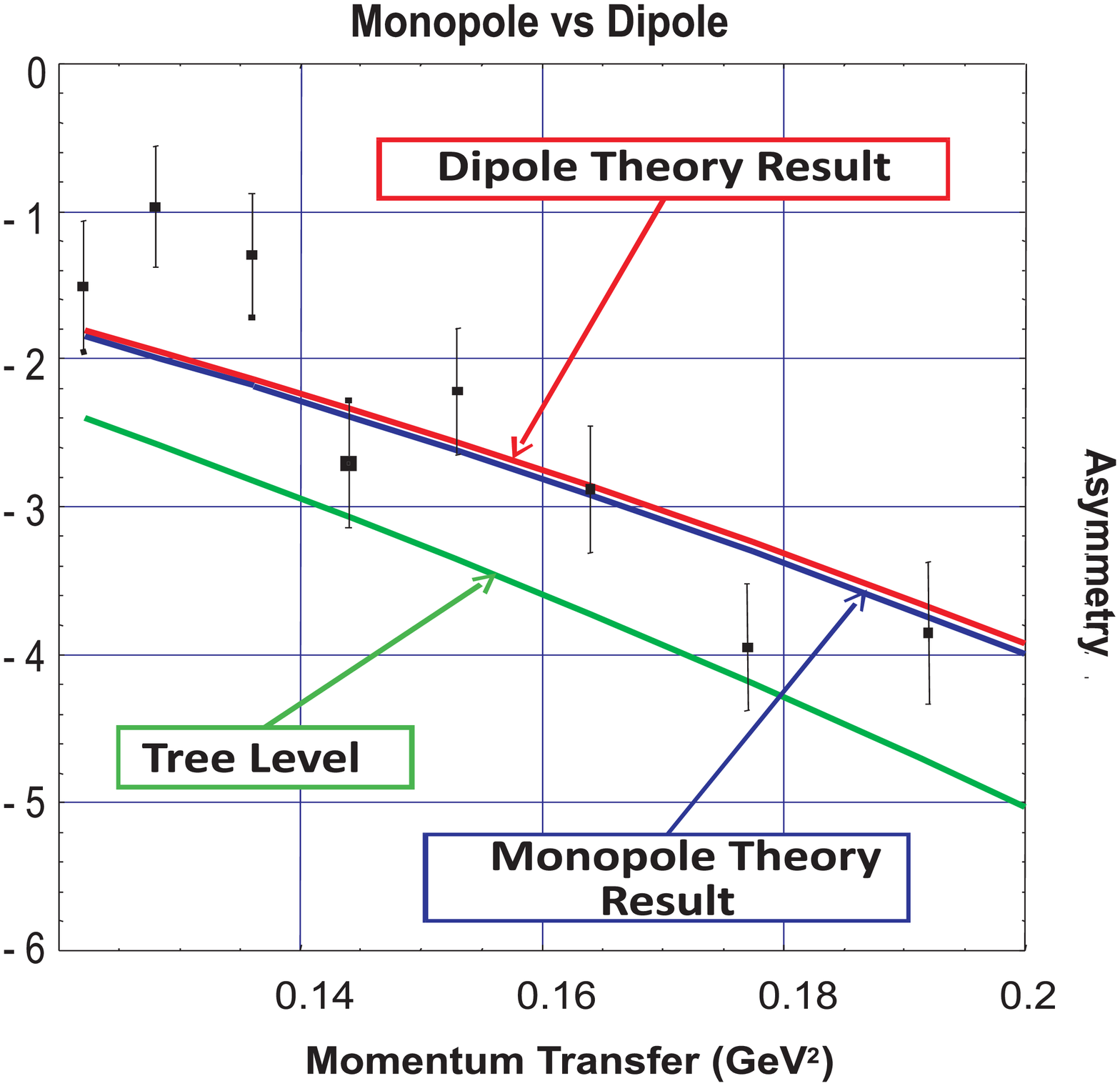}\includegraphics[scale=0.25]{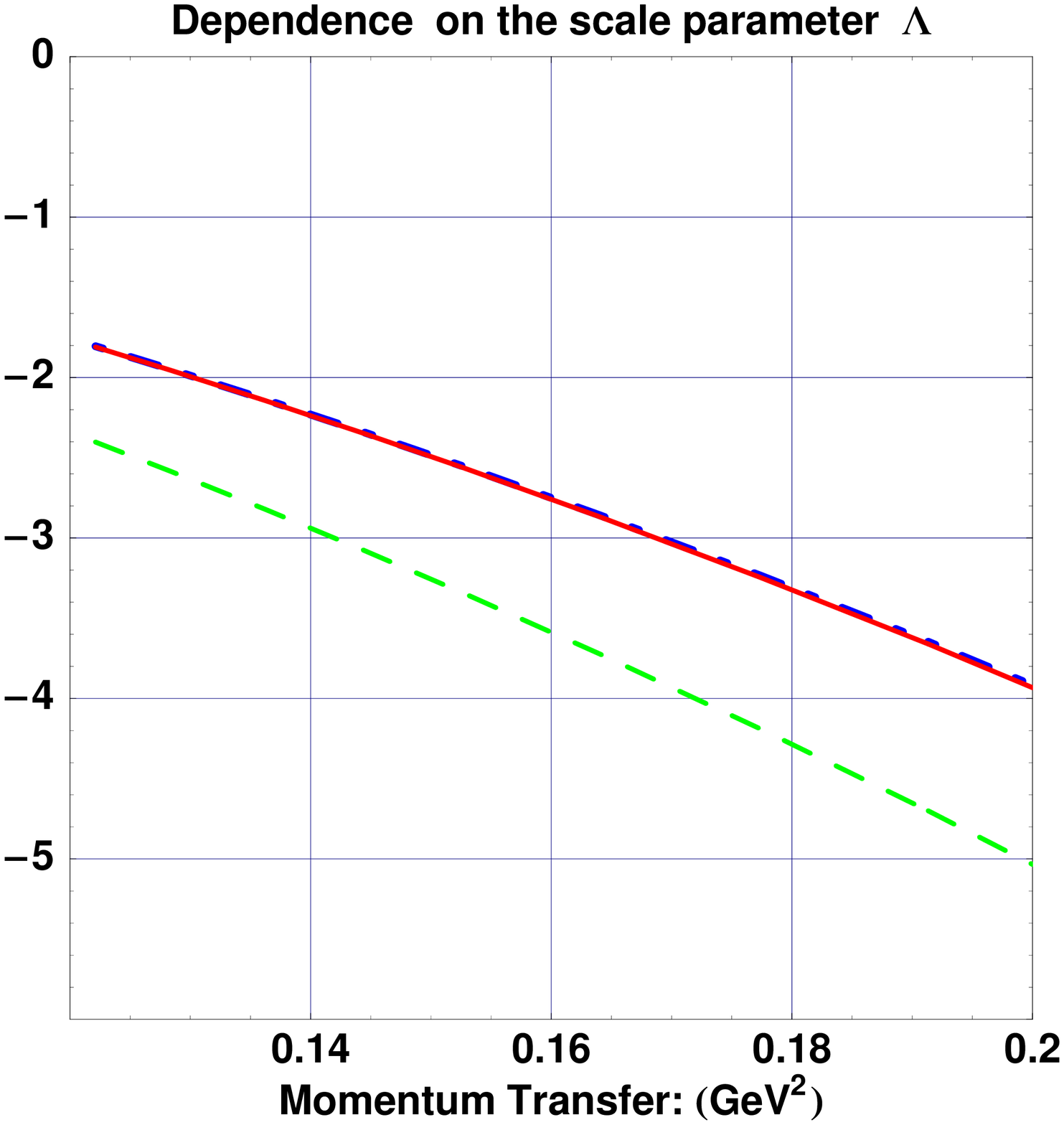}
\par\end{centering}

\caption{a) Asymmetry for the PV electron-proton scattering is given on the left figure. Dipole theory
results are given by LO+NLO contributions, and $A_{NVS}$ is no-vector-strange
asymmetry given in \cite{G0}. 
b) Figure in the centre shows the difference in the asymmetries for monopole and dipole formfactors. 
c) Figure on the right shows the difference in asymmetries for the two
limiting values of scale parameter, $\Lambda^{2}=0.4\cdot m_{N}^{2}$(solid
red curve) and $\Lambda^{2}=2.0\cdot m_{N}^{2}$(dashed blue curve).
Dashed green curve is a leading order contribution. }

\end{figure}
It is clear from Fig.{[}1a{]} that NLO PV effects are of the order
of 20\%. Also we can see that our results are in excellent agreement
with theoretical predictions of G0 group \cite{G0}. Comparing our predictions
with the G0 \cite{G0} experimental asymmetry we can conclude that
there is an evident discrepancy between theory and experiment. Recalling
the fact that our calculations were completed without an account for
the sea of the strange quarks in the nucleon, good explanation to
this would be a non negligible strange content of the nucleon. Moreover,
we have used a model dependent form-factors and it is reasonable to
investigate impact of this model dependence on calculated asymmetries.
In this case we looked at the difference between monopole and dipole
results and dependence on the scale parameter $\Lambda$ (see Fig.{[}1b,c{]}).
From Fig.{[}1b,c{]} it is evident that calculations of the asymmetries
are independent of the choice of the type of form-factors and have
virtually no dependence on the scale parameter $\Lambda$. Although
model dependence will become evident when calculating absolute cross
sections, for the asymmetries calculations presented results are model
independent. We also reserved the kinematic dependence in all types
of our radiative corrections. It will make it easier to adopt our
results to the current and future parity-violating experiments for
any lepton-hadron scattering processes.

\section{Hadronic Sector }

Tremendous success of the ChPTh in the description of the hadronic
interactions in the non-perturbative regime of the QCD attracted attention
of the physics community for decades. To calculate amplitudes or cross
sections, we need a theoretical input at the NLO retaining full kinematic
dependencies. To date, there are several packages
(FeynArts, FormCalc, FeynCalc and Form) allowing us to produce semi-automatic
calculations in the high energy physics. Although FeynArts and FormCalc
were originally designed for SM calculations, the flexibility
of the programs allows us to extend them to interactions appropriate
for the hadronic sector. This was a main reason to use FeynArts and
FormCalc as a base languages for the automatization of chiral hadronic
calculations. Here we have used a CHM which detailed description reader can find in \cite{CHM}. 
As an application and test of this model, we decided to extract strong coupling constants of ChPTh
from the experimental values of BR to decays of resonances to baryons,
such as: $\Sigma^{*+}\rightarrow\Sigma^{0}+\pi^{+},$ $\Sigma^{*-}\rightarrow\Lambda+\pi^{+},$
$\Xi^{*0}\rightarrow\Xi^{*-}+\pi^{+}$ and $\Delta^{++}\rightarrow p+\pi^{+}$.
The BR are centered about 1.4, 1.4, 1.3 and 1.5 respectively and as it
is expected, leading order calculations are SU(3) conserving and will
result in equally valued BR. This fact prompts us to look at the NLO
corrections. The NLO calculations were completed with an account of
the octet of mesons, baryons and decuplet of resonances participating
in the one loop diagrams. If we take experimental BR and fit strong
coupling constants C and H to the values of BR we can reproduce results
predicted earlier in \cite{MB}. For the each decay outlined earlier
in this article, parametric plots for the C coupling constant as a
function H are shown in Fig.{[}2{]}. 

\begin{figure}
\centering{}\includegraphics[scale=0.36]{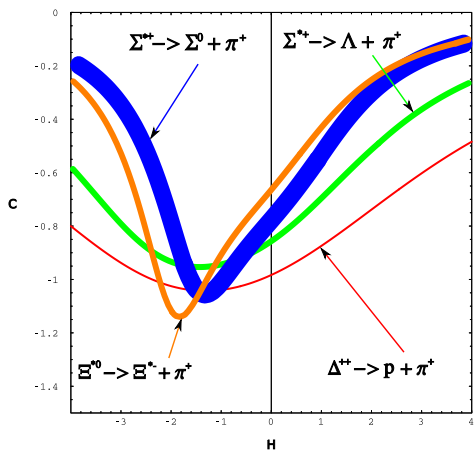}\caption{Parametric plots for the strong coupling constants C and H fitted
to the BR of $\Sigma^{*+}\rightarrow\Sigma^{0}+\pi^{+},$ $\Sigma^{*-}\rightarrow\Lambda+\pi^{+},$
$\Xi^{*0}\rightarrow\Xi^{*-}+\pi^{+}$ and $\Delta^{++}\rightarrow p+\pi^{+}$.
Thickness of the curves represents the experimental uncertainty taken
from the BR for these decays.}

\end{figure}
From Fig.{[}2{]} it is clear that all curves indicate the central value of
the C coupling constant is around $|C|=1.0\pm0.2$. From this we can
derive constrains on the H coupling constant as $H=-1.5\pm0.5$. Our predictions are consistent with the results of \cite{MB}:
$|C|=1.2\pm0.1$ and $H=-2.1\pm0.7$. In general, these calculations served us
as an excellent test of rather involved computer based calculations.
The results of the CHM can be expressed analytically, using FormCalc
with an amplitude expressed in the Passarino-Veltman basis as well
as numerically with application of the package LoopTools. The future
applications of the CHM can be seen in the automatization of the calculations
of NLO radiative corrections for the production/decay channels of
hadronic physics. 

\section{Acknowledgment}

This work has been supported by an NSERC Discovery Grant.

\end{document}